\documentclass{article}
\usepackage[utf8]{inputenc}
\usepackage{mathtools}
\usepackage[table]{xcolor}
\usepackage{amssymb,graphicx}
\usepackage{epstopdf}
\usepackage{amsmath,amsfonts}
\usepackage{epsfig}
\usepackage{xcolor}
\usepackage[T1, T2A]{fontenc}
\usepackage[english]{babel}
\usepackage{float}
\usepackage{soul}
\usepackage{graphicx}
\usepackage{subcaption}
\usepackage{geometry}
\usepackage{hyperref}
\usepackage{psfrag}
\usepackage{slashed}
\usepackage{braket}
\usepackage{bm}
\usepackage{siunitx}
\usepackage{soul}
\begin{document}
\begin{flushright}
INR-TH-2023-019
\end{flushright}
\vspace{10pt}
\begin{center}
  {\LARGE \bf K-inflation: the legitimacy of classical treatment} \\
\vspace{20pt}
Y.~Ageeva$^{a,b,}$\footnote[1]{{\bf email:}
    ageeva@inr.ac.ru}, P. Petrov$^{a,}$\footnote[2]{{\bf email:}
    petrov@inr.ac.ru}\\
\vspace{15pt}
  $^a$\textit{
Institute for Nuclear Research of
         the Russian Academy of Sciences,\\  60th October Anniversary
  Prospect, 7a, 117312 Moscow, Russia}\\
\vspace{5pt}
$^b$\textit{
  Institute for Theoretical and Mathematical Physics,
  M.V.~Lomonosov Moscow State University,\\ Leninskie Gory 1,
119991 Moscow,
Russia
}
    \end{center}
    \vspace{5pt}
\begin{abstract}
%
  In this paper we consider a general theory of k-inlation and find out, that it may be in strong coupling regime. We derive accurate conditions of classical description validity using unitarity bounds for this model. Next, we choose simple toy model of k-inflation and obtain the explicit condition, which guarantees that the generation of perturbations is performed in a controllable way, i.e the exit from the effective horizon occurs in the weak coupling regime. However, for the same toy model the corresponding experimental bounds on a non-linear parameter $f^{\text{equil}}_{\text{NL}}$ associated with non-Gaussianities of the curvature perturbation provide much stronger constraint than strong coupling absence condition. Nevertheless, for other known models of inflation this may not be the case. Generally, one should always check if classical description is legitimate for chosen models of inflation.
\end{abstract}

\section{Introduction}
\label{sec:intro}
Nowadays, inflation \cite{Starobinsky:1980te,Guth:1980zm,Sato:1980yn, Linde:1981mu} is a very successful paradigm for understanding the properties of the early Universe. Among many models of inflation, we choose k-inflation model \cite{Armendariz-Picon:1999hyi,Garriga:1999vw} for our purposes. In such models the lagrangian involves non-canonical kinetic term, which drives the cosmological evolution. Although k-inflation theories are known as free from obvious pathologies, we address to the examination of \textit{strong coupling problem} in k-inflation. The energy scale of strong coupling is an important parameter in an effective QFT. In other words, it is the maximum energy below which the effective QFT description is valid. The strong coupling energy scale can often be qualitatively estimated by naive dimensional analysis, see for example \cite{Ageeva:2018lko,Ageeva:2020gti,Ageeva:2020buc,Ageeva:2021yik}. There are different notations of strong coupling concern in other works, for example, in Ref.~\cite{Baumann:2011dt} authors use the relation $\frac{\mathcal{L}_3}{\mathcal{L}_2}$, so that when this ratio becomes larger than one, the theory is strongly coupled. However, in this paper we want to obtain more accurate estimations of strong coupling, thus we stick to another criterion given in Refs.~\cite{Ageeva:2022fyq,Ageeva:2022nbw,Grojean:2007zz}. Our notation of legitimacy of classical treatment takes into account precise estimates using \textit{unitarity bounds} that follow from general unitarity relations which must be used in order to proceed to the precise analysis of mentioned problem. 

We show -- firstly by the preliminary analysis of cubic order action for scalar perturbation -- that strong coupling problem indeed arises in the most general setup for k-inflation.  The simple estimation of $s$-channel matrix element for the $2\to2$ scattering processes  and applied unitarity bound, provide some non-trivial conditions on model functions and parameters. Next, in order to improve our estimation, we turn to the explicit calculation of all channels: $s-$, $t-$, and $u-$matrix elements. The structure of k-inflation model lagrangian and cubic order action for scalar perturbation in this model lead to non-trivial cancellations in the final answer for matrix element. It also turned out that $t$- and $u$-elements are suppressed compared to $s$-channel and the factor of suppression is the slow roll parameter $\epsilon$, which is usually a small quantity during inflation $\epsilon\ll 1$. Using more accurate result for $s$-channel element, unitarity bound provides final constraint on the parameters of the model. 

Our next step is to choose a simple toy model of k-inflation in order to show how to apply the  unitarity constraints. It turns out that the latter gives the lower bound on the slow-roll parameter $\epsilon$. 

As we mention above, the analysis of strong coupling involves cubic order action for scalar perturbation. The same expansion is used in the calculations of \textit{non-Gaussianity of the curvature perturbation}. Thus, it is interesting to compare the conditions on the parameters of the model that comes from observational bound on non-Gaussianity \cite{Planck:2019kim} and from the validity of classical description. Note that these bounds have different nature: conditions from non-Gaussianity are experimental constraints, while the strong coupling absence guarantees that our classical description is legitimate during considered times. 

Again, working with a specific toy model of k-inflation, the non-Gaussianities also lead to the lower bound on the slow roll parameter $\epsilon$. However, the condition from non-Gaussianities turns out to be much stronger than the bound from strong coupling analysis for the chosen toy model of k-inflation. We emphasize that this result is obtained for the specific model of k-inflation: the situation may differ in other models of inflation. In other words, one should check if the classical description is legitimate for the chosen theory. For example, other models of inflation may not lead to the cancellations in cubic order action between the leading terms, so this can make the conditions of  strong coupling absence more restrictive.

This paper is organized as follows: a brief review of general k-inflation model is given in Sec.~\ref{sec:model}. Then the analysis of the strong coupling problem is addressed in Sec.~\ref{sec:StrongCoupling}: simple estimations of $s$-channel matrix element and naive condition from unitarity bound are given in subsection \ref{subsec:naive}. This allows us to highlight the terms which provide the strongest constraints. More accurate calculations of $s$-, $t$-, $u$- channels matrix elements and the final condition from unitarity bound are given in subsection~\ref{subsec:diagram}. The Sec.~\ref{sec:nongauss} dedicated to short discussion of the formulas for a non-linear parameter $f^{\text{equil}}_{\text{NL}}$ associated with non-Gaussianities of the curvature perturbation. Finally, in Sec.~\ref{sec:concrete_model} we stick to the specific simple model of k-inflation, find corresponding constraints on model parameter from strong coupling analysis as well as from bounds for non-Gaussianities. The paper ends with the conclusion in Sec.~\ref{sec:conclusion}. The Appendix~\ref{app:second_order} is dedicated to the derivation of second order action for scalar perturbation. We explicitly show, that at considered times second order action has a canonical form due to the smallness of slow-roll parameter as well as due to the assumption, that we work at such energies that much bigger than classical energy scales. Thus the standard calculations of scattering amplitudes with the use of energy-momentum conservation laws are acceptable. The Appendix~\ref{app:lambdas} collects full expressions of the couplings from cubic order action for scalar perturbation.  The Appendices~\ref{app:t_u_ch} and~\ref{app:s_ch} contain the general clarifications to the calculations of $s$-, $t$-, and $u$-channel matrix elements from the subsection~\ref{subsec:diagram} as well as the discussion of one interesting subtlety which arises in the calculation of $s$-channel matrix element from the same subsection~\ref{subsec:diagram}.

\section{Generalities}
\label{sec:model}

In this paper we 
consider a class of k-inflation models in the framework of the following action:
\begin{equation}
        \label{action}
        \mathcal{S} = \int d^3x dt  \sqrt{-g}\mathcal{L},
    \end{equation}
where $\sqrt{-g} \equiv \sqrt{\gamma}$ with three dimensional metric tensor and the determinant $\gamma \equiv \text{det}(^{(3)}\gamma_{ij})$; the lagrangian reads
    \begin{eqnarray}
    \cal L&=&G_2(\phi, X)+ \frac{M_{Pl}^2}{2}R, 
    \label{Hor_L}\\
        X &=& -\frac{1}{2}g^{\mu\nu}\partial_{\mu}\phi\partial_{\nu}\phi,
    \nonumber
    \end{eqnarray}
and $G_2(\phi,X)$ is an arbitrary function of scalar field and its kinetic term;  $R$ is the Ricci scalar. Here we also note that we work
in the Einstein frame through the whole paper. Further we will use the metric
signature as $(-,+,+,+)$.

We consider the flat FLRW space-time with a scale factor $a(t)$, where $t$ is the cosmic time, so the background equations read \cite{DeFelice:2011zh} 
\begin{subequations}
\label{EoM}
\begin{align}
    3M_{Pl}^2H^2+G_2-2XG_{2X}=0,\\
    3M_{Pl}^2H^2+2M_{Pl}^2\dot{H}+G_2=0,
\end{align}
\end{subequations}
where $H = \dot{a}/a$ is the Hubble parameter.
As it was pointed out in \cite{DeFelice:2011zh,Seery:2005wm,Chen:2006nt}, one can obtain k-inflation cosmology solving these equations for the specific form of $G_2$ function. 
The inflation occurs in the slow roll regime, i.e. $\epsilon\ll 1$ \cite{Weinberg:2008zzc}, where $\epsilon$ is a standard slow-roll parameter which is given by:
\begin{equation}
\label{eps}
    \epsilon\equiv -\frac{\dot{H}}{H^2} = \frac{X G_{2X}}{M_{Pl}^2H^2}\;.
\end{equation}
The condition $\epsilon\ll 1$ can be satisfied with some specific choice of $G_2$ form. For instance, one can choose $G_2$ as \cite{Armendariz-Picon:1999hyi,Peng:2016yvb}
\begin{align}
\label{specific_G2_K_L}
    G_2(\phi,X) = K(\phi)X + L(\phi)X^2\;,
\end{align}
where the dimensions of the functions $K(\phi)$, $L(\phi)$, and $X$ are as follows $[K] = 2$, $[L] = 0$, and $[X] = 2$; moreover, we note that in our setup we have $[\phi] = 0$. This form of $G_2$ indeed admits the slow-roll inflation solution, and a necessary condition for the accelerated expansion in this case reads \cite{Peng:2016yvb}:
\begin{equation*}
    \frac{X(K+2XL)}{M_{Pl}^2H^2}\ll 1.
\end{equation*}
To obtain the latter we also use an expression for the Hubble parameter during inflation (up to the leading order by $\epsilon$)\cite{Peng:2016yvb}:
\begin{equation*}
    H^2 \approx -\frac{G_2}{3M_{Pl}^2}.
\end{equation*}
        
In order to explore the stability of the model, the strong coupling problem as well as to calculate the primordial scalar non-Gaussianities we need to expand the action \eqref{action} up to the second and the third order in the perturbations. In this paper we concentrate on the scalar sector of perturbations only, since  this sector usually provides the strongest conditions; for instance, see Ref. \cite{Ageeva:2022asq}. Later, when we turn to the concrete model of k-inflation, we prove that scalar sector indeed gives the strongest constraints. To this end, considering the perturbations about some background solution, we choose the following form of the metric \cite{DeFelice:2011zh}: 
\begin{align*}
    &ds^2 = -[(1+\alpha)^2 - a^{-2} \text{e}^{-2\mathcal{R}}(\partial\beta)^2]dt^2 \nonumber\\
    &+ 2\partial_i\beta dt dx^i + a^2 \text{e}^{2\mathcal{R}} d{\bf x}^2 \; ,
\end{align*}
where  $\alpha $ and $\beta$ are non-dynamical scalar perturbations, while $\mathcal{R}$ is a physical one. We also note, that we work with the \textit{unitary} gauge, i.e. $\delta\phi = 0$, which fixes the time-component of a gauge-transformation vector, see \cite{DeFelice:2011zh,Ageeva:2022asq,Kobayashi:2019hrl} for the details. Solving the constraints for $\alpha$ and $\beta$, we write  the unconstrained action for
scalar perturbation $\mathcal{R}$ \cite{DeFelice:2011zh,Kobayashi:2015gga}
\begin{equation}
    \label{second}
    \mathcal{S}^{(2)}_{\mathcal{R}\mathcal{R}} = \int dt\; a^3 d^3x \; \mathcal{G}_S \left(\dot{\mathcal{R}}^2 - \frac{c_S^2}{a^2}(\vec{\nabla} \mathcal{R})^2\right),
\end{equation}
 where
\begin{equation}
\label{Gs}
    \mathcal{G}_S = \frac{XG_{2X}+2X^2G_{2XX}}{H^2}= \frac{\Sigma}{H^2},
\end{equation}
where $\Sigma \equiv XG_{2X}+2X^2 G_{2XX}$; next
\begin{equation}
    \label{cs}
    c_S^2 = \frac{M_{Pl}^2H^2\epsilon}{XG_{2X}+2X^2G_{2XX}} = \frac{M_{Pl}^2H^2\epsilon}{\Sigma}.
\end{equation}
Using the expression \eqref{cs}, we can rewrite formula \eqref{Gs} as  
\begin{equation}
\label{Gs_cs_eps}
    \mathcal{G}_S  = M_{Pl}^2\frac{\epsilon}{c_S^2},
\end{equation}
where the ratio $\frac{\epsilon}{c_S^2}$ generally is not small. 

Briefly turning to the stability analysis, we require that 
\begin{align}
        \mathcal{ G}_S>0, \; c_S^2 >0,
        \label{stability_conditions}
    \end{align}
to avoid ghost and gradient instabilities
as well as we require that the speed of perturbations  does 
not exceed the  speed of light,
\begin{align*}
    c_S^2\leq 1 \; .
\end{align*}
The latter condition is  necessary for  
the existence of the UV  completion, see 
\cite{Adams:2006sv,deRham:2013hsa} for the details.

\section{Strong coupling regime in k-inflation model}
\label{sec:StrongCoupling}

This Section is dedicated to the computation of the unitarity bounds and corresponding constraints on the parameters of the model. We remind, that we consider pure scalar sector and we take into account only cubic order expansion of the action \eqref{action} by the scalar perturbation $\mathcal{R}$. This Section consists of two parts: in the first part we estimate which terms from cubic order action provide the leading contributions to unitarity bound, while in the second part we use these leading terms in order to proceed to the accurate calculation of the corresponding matrix elements and final conditions for the validity of the classical description.

\subsection{\textit{Preliminary analysis}}
\label{subsec:naive}
In order to show, that we indeed face the strong coupling regime in the considered class of k-inflation model \eqref{action}, let us firstly carry out the simple dimensional analysis of noted problem. To this end we write the full unconstrained cubic order action for scalar perturbation $\mathcal{R}$ \cite{Ageeva:2020gti}:
 \begin{align}
 \label{cubic}
        \mathcal{S}^{(3)}_{\mathcal{R}\mathcal{R}\mathcal{R}}&=   \int dt\text{ }a^3d^3x \left\{ \Lambda_1 \dot{\mathcal{R}}^3 
        + \Lambda_2 \dot{\mathcal{R}}^2\mathcal{R} 
        + \Lambda_3 \dot{\mathcal{R}}^2 \frac{\partial^2 \mathcal{R}}{a^2}  
        +\Lambda_4 \dot{\mathcal{R}}\mathcal{R} \frac{\partial^2 \mathcal{R}}{a^2}\right. \nonumber \\
        &  + \Lambda_5 \dot{\mathcal{R}} \frac{\left(\partial_i \mathcal{R} \right)^2}{a^2} 
        +\Lambda_6 \mathcal{R} \frac{\left(\partial_i \mathcal{R} \right)^2}{a^2} 
        + \Lambda_7 \dot{\mathcal{R}} \frac{\left(\partial^2 \mathcal{R} \right)^2}{a^4} 
        + \Lambda_8 \mathcal{R} \frac{\left(\partial^2 \mathcal{R} \right)^2}{a^4} 
        + \Lambda_9 \frac{\partial^2 \mathcal{R} \left(\partial_i \mathcal{R} \right)^2}{a^4} 
        \nonumber \\
        &+ \Lambda_{10} \dot{\mathcal{R}} \frac{\left(\partial_i \partial_j \mathcal{R} \right)^2}{a^4} 
        + \Lambda_{11} \mathcal{R} \frac{\left(\partial_i \partial_j \mathcal{R} \right)^2}{a^4} 
        + \Lambda_{12} \dot{\mathcal{R}} \partial_i \mathcal{R} \partial_i \psi 
        + \Lambda_{13} \frac{\partial^2 \mathcal{R} \partial_i \mathcal{R} \partial_i \psi}{a^2} 
        +  \Lambda_{14} \dot{\mathcal{R}} \left(\partial_i \partial_j \psi \right)^2 
        \nonumber \\
        &+ \Lambda_{15} \mathcal{R} \left(\partial_i \partial_j \psi \right)^2
        + \left.
        \Lambda_{16} \dot{\mathcal{R}} \frac{\partial_i \partial_j \mathcal{R} \partial_i \partial_j \psi}{a^2}
        + \Lambda_{17} \mathcal{R} \frac{\partial_i \partial_j \mathcal{R} \partial_i \partial_j \psi}{a^2} \right\},
    \end{align}
where $\partial^2 = \partial_i\partial_i$ and  
    \begin{equation}
    \label{psi}
        \psi=\partial^{-2}\dot{\mathcal{R}}.
    \end{equation}
Actually, there are non-trivial cancellations in the models with the lagrangian \eqref{Hor_L} among $ \Lambda_7, \ldots,  \Lambda_{11}$ terms from the action \eqref{cubic}. Indeed, substituting the lagrangian \eqref{Hor_L}, as well as expressions \eqref{Gs} and \eqref{cs} into the general formulas for these coefficients, which are listed in Appendix~\ref{app:lambdas}\footnote{All other couplings $\Lambda_1$--$\Lambda_6$, $\Lambda_{12}$--$\Lambda_{17}$ expressions are listed in Appendix~\ref{app:lambdas} as well.},  we arrive to
\begin{align}
\label{l7-l11}
        &\Lambda_7  
        =\frac{M_{Pl}^2}{2H^3}, \quad \Lambda_8=
        -\frac{3M_{Pl}^2}{2H^2}, \quad \Lambda_9=
        -\frac{2M_{Pl}^2}{H^2},  \nonumber\\
        &\Lambda_{10}=
        -\frac{M_{Pl}^2}{2H^3},  \quad \Lambda_{11}=\frac{3M_{Pl}^2}{2H^2},
\end{align}
and after quite simple integration by parts \cite{DeFelice:2011zh} this part of cubic action significantly simplifies as follows
\begin{align*}
       &\mathcal{S}^{(3)}_{7,8,9,10,11}=\int  dt \;  d^3x \; \frac{1}{a} \; 
         \Big\{ 
         \Lambda_9 \partial^2 \mathcal{R} \left(\partial_i \mathcal{R} \right)^2 
        + (\Lambda_{10} \dot{\mathcal{R}}+\Lambda_{11} \mathcal{R}) \Big(\left(\partial_i \partial_j \mathcal{R} \right)^2 - \left(\partial^2 \mathcal{R} \right)^2\Big)
        \Big\} \\
        &= \int  dt \;  d^3x \; 
         \left\{\frac{d}{dt}\left(\frac{\Lambda_{10}}{3a}\right)-\frac{\Lambda_{11}}{a}-\frac{2}{3a}\Lambda_9\right\} \mathcal{R} \Big( \left(\partial^2 \mathcal{R} \right)^2-\left(\partial_i \partial_j \mathcal{R} \right)^2\Big)
    ,
    \end{align*}
where curly brackets read
\begin{align}
\label{non_zero}
    &-\frac{1}{a}\left(H\left(\frac{\Lambda_{10}}{3}\right)-\frac{d}{dt}\left(\frac{\Lambda_{10}}{3}\right)+\Lambda_{11}+\frac{2}{3}\Lambda_9\right) \nonumber\\
    &= -\frac{M_{Pl}^2\epsilon}{2aH^2},
\end{align}
where non-zero contribution comes from the second term with the time derivative, i.e. from $\frac{d}{adt}\left(\frac{\Lambda_{10}}{3}\right)$, while the combination of other three terms with $\Lambda_{9},\Lambda_{10},\Lambda_{11}$ gives zero. After that, we will denote this contribution from formula \eqref{non_zero} as
\begin{equation}
     \label{L_7891011}
\Lambda_{*} \equiv  -\frac{M_{Pl}^2\epsilon}{2H^2}.
\end{equation}

To find the conditions of the validity of the classical description, we turn to the generalized  unitarity bound and use the method which was described in \cite{Ageeva:2022nbw}. According to this method, firstly we need to rewrite the quadratic action \eqref{second} in the following canonical form:
\begin{align}
\label{second_canonic}
    \mathcal{S}^{(2)}_{\mathcal{R}\mathcal{R}} =  \frac{1}{2}\int d^3x d\eta \left[\tilde{\mathcal{R}}^{\prime\; 2} -c_S^2 (\vec{\nabla}\tilde{\mathcal{R}})^2  \right],
\end{align} 
where we use quite familiar notation of Mukhanov-Sasaki variable 
 $\tilde{\mathcal{R}} = z \mathcal{R}$ with $z = a\sqrt{2\mathcal{G}_S}$. Here we also have
$d\eta = \frac{dt}{a}$ as a conformal time, which we will use in the calculations below; the prime means the derivative with respect to the conformal time $' \equiv \frac{d}{d\eta}$. In Appendix \ref{app:second_order} we show, how we obtain the canonical form \eqref{second_canonic}. We also note, that in this work we study only such models, which involves $c_S$ being a constant in the leading order by slow-roll parameter $\epsilon$. Thus, eq. (16) is a standard time-independent oscillator action. Next, in Appendix \ref{app:lambdas} we write down the cubic order action \eqref{cubic} in terms of $\tilde{\mathcal{R}}$, see formula \eqref{app:cubic_canonic}. Having the latter, we can proceed to 
the analysis of the potential strong coupling problem. To this end, making use of all terms in the cubic action \eqref{app:cubic_canonic},
with $\Lambda_i$ replaced by $\Lambda_{i,(j)} \propto
\frac{\Lambda_i}{\mathcal{G}_S^{3/2}}$, it is straightforward to estimate $2\to2$ scattering amplitude, while in the subsection \ref{subsec:diagram} we calculate this amplitude accurately\footnote{New index $(j)$ can be explained by the replacement of $\mathcal{R}$ to $\tilde{\mathcal{R}}/z$, where $z$ depends on conformal time, so taking the derivative with respect to conformal time provide several terms with different $\Lambda_{i,(j)}$, see Appendix \ref{app:lambdas} for details.}. Firstly, the dimensional analysis leads us to the schematic formula for the tree $2\to2$ matrix element\footnote{For this kind of estimations we  consider the s-channel matrix element only.} \cite{Ageeva:2022asq} 
\begin{align}
\label{M_naive}
    M_{i,(j)} \sim \frac{1}{E^2}\cdot \Big\{\Lambda_{i,(j)}\cdot  E^{a}\cdot  \Big(\frac{E}{c_S}\Big)^{b}\Big\}^2,
\end{align}
where $a$ and $b$ are the number of time and spatial derivatives for each term in \eqref{app:cubic_canonic}. We consider the center-of-mass frame for our purposes. The conservation laws for the latter are as follows
\begin{subequations}
\label{conservation}
\begin{align}
    \label{law_momenta}
    &\vec{p}_1 + \vec{p}_2 = \vec{p}_3 + \vec{p}_4 = 0,\\
    &E_1 + E_2 = E_3 + E_4 = E,\\
    &|\vec{p}_1| = |\vec{p}_2|,\quad |\vec{p}_3| = |\vec{p}_4|,
\end{align}
where $\vec{p}_{1,2}$, $E_{1,2}$ and $\vec{p}_{3,4}$, $E_{3,4}$ are the incoming and outcoming particles momenta and energies, respectively. Surely, the use of such conservation laws at the considered early times are legitimate, since the action for perturbations has the form of \eqref{second_canonic}. Next, we find
\begin{equation}
    E_{1,2,3,4} = \frac{E}{2},
\end{equation}
where $E$ is the center-of-mass energy and we note that $E$ is a conformal energy.
Due to \eqref{second_canonic}, the dispersion relation reads
\begin{equation}
    E_{1,2,3,4} = c_S p_{1,2,3,4},
\end{equation}
thus
\begin{equation}
     p_{1,2,3,4} = \frac{E}{2c_S}.
\end{equation}
\end{subequations}
Coming back to the formula \eqref{M_naive}, the factor $\frac{1}{E^2}$ presents the s-channel propagator. Next, since the energy and momentum of the scalar are related by $\omega = c_S p$ (note, that we reserve the
notation $E$ for the center-of-mass energy), spatial momentum of an incoming or outgoing
scalar is of order $p \sim E/c_S$. This clarifies the factor $\Big(\frac{E}{c_S}\Big)^{b}$, coming from the Fourier of spatial derivative. Moreover, in the case of center-of-mass frame the energies of incoming (noted as $\omega_{1,2}$) and outgoing (noted as $\omega_{3,4}$) scalars are $\omega_{1,2,3,4} \sim E$, thus we count the possible factor $E^a$ from the Fourier of time derivative. We square the expression in curve brackets in eq. \eqref{M_naive} since for our naive estimations we consider the easier case when both vertices are the same. The corresponding partial wave amplitude (PWA) is given by \cite{Ageeva:2022nbw,Ageeva:2022asq,Grojean:2007zz}
\begin{equation}
\label{PWA}
    \tilde{a}_l = \frac{1}{2c_S^3}\frac{1}{32\pi}\int d(cosx)P_l(cosx) M,
\end{equation}
so, omitting all numerical coefficients we can write for $l=0$ and for each $M_{i,(j)}$
\begin{equation}
\label{a0_naive}
    (\tilde{a}_0)_{i,(j)} \sim \frac{M_{i,(j)}}{c_S^3}. 
\end{equation} 
It is known from Refs. \cite{Ageeva:2022nbw,Grojean:2007zz,Ageeva:2022asq} that the amplitudes at classical energy scales saturate the unitarity bound. 
The classical energy scale is given by Hubble parameter $H$, and the  latter was obtained in cosmic time $t$, see eqs.~\eqref{EoM}. However, the amplitudes \eqref{a0_naive} are given in conformal time $\eta$, thus, if we are supposed to compare strong coupling energy scales with classic one, we must 
use $\mathcal{E} = E/a$, where $\mathcal{E}$ is corresponding energy in cosmic time $t$.  
Finally, the unitarity bound $|\tilde{a}_0| = 1/2$ provides the set of different scales $\mathcal{E}$ from each matrix element $M_{i,(j)}$, and the condition of the strong coupling absence $\mathcal{E} \gg H$ provides the set of the following constraints \footnote{Some of the amplitudes provide the same constraint. Moreover, the terms in cubic action with the dimension of coupling $[\Lambda_{i,(j)}] \geq 0$ do not provide any conditions of strong coupling absence.}: 
\begin{subequations}
\label{inequalities}
\begin{align}
    \frac{1}{\epsilon^{3/2}}\ll \frac{H^3M_{Pl}^7}{\Sigma^{5/2}},\\
    \epsilon^{1/2} \ll \frac{\Sigma^{3/2}}{H^5 M_{Pl}},\\
    \frac{1}{\epsilon^{3/2}}\ll \frac{M_{Pl}^3 \Sigma^{3/2}}{H \lambda_1^2},\\
    \frac{1}{\epsilon^{3/2}}\ll \frac{M_{Pl}^3}{H  \Sigma^{1/2}},\\
    \label{str1}
    \frac{1}{\epsilon^{7/2}}\ll \frac{H^3 M_{Pl}^7}{\Sigma^{5/2}},\\
    \label{str2}
    \frac{1}{\epsilon^{7/2}}\ll \frac{ M_{Pl}^3}{H\Sigma^{1/2}},
\end{align}
\end{subequations}
where $\lambda_1 \equiv X^2 G_{2XX} + X^3 G_{2XXX}/3$ 
and we put all related calculations in Appendix \ref{app:lambdas}. Since $\epsilon^{-1}$ is an enhancement factor, the strongest conditions are \eqref{str1} and \eqref{str2}, coming from $\Lambda_3-\Lambda_{6}, \Lambda_{*}, \Lambda_{13}, \Lambda_{16}, \Lambda_{17}$ terms. We will not consider the terms with other couplings in our more accurate analysis of the amplitudes since they provide suppressed contribution.

\subsection{\textit{Strong coupling problem: accurate analysis}}
\label{subsec:diagram}
\begin{figure}[H]
    \centering
\includegraphics[scale=0.3]{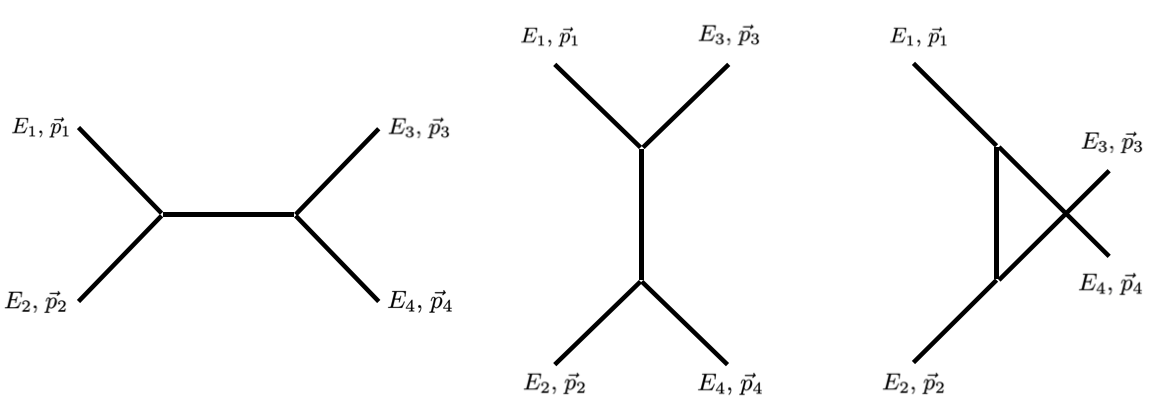}
    \caption{Tree level diagrams for $2\to 2$ process: $s$-, $t$-, and $u$-channels, respectively.}
    \label{fig:channels}
\end{figure}

In this subsection we go ahead to precisely calculate tree matrix elements -- $s$-, $t$-, and $u$-channels -- and find more accurate constraints on model parameters from the strong coupling problem analysis.
We mention once again, that we work with the center-of-mass frame. We start with $s$-channel, corresponding diagram is shown in Fig. \ref{fig:channels} (left one) and corresponding conservation laws are given by eqs.~\eqref{conservation}. Thus, the $s$-channel matrix element is
\begin{align}
\label{s-channel-acc}
    iM_s = - \frac{i (E^6 \Sigma^2 +  E^4 \Sigma (-8 H^2 M_{Pl}^2+5\Sigma)a^2H^2 + 4 E^2(-2H^2 M_{Pl}^2+\Sigma)^2 a^4H^4)}{128\epsilon^2\Sigma M_{Pl}^4a^6H^6}.
\end{align}
Next, the expression for the $t$-channel matrix element (corresponding diagram is given in Fig. \ref{fig:channels}, central one) reads
\begin{align}
\label{t-channel-acc}
	iM_t = - \frac{i \Big\{E^3(x^2-1) + 8Ea^2H^2\big(3+2x-4x^2+ \frac{2H^2 M_{Pl}^2(x-2)}{\Sigma}\big)\Big\}^2}{1024 \epsilon M_{Pl}^2a^6H^4(x-1)},
\end{align} 
where $x \equiv \text{cos}\;\theta$, where $\theta$ is an angle between $\vec{p}_1$ and  $\vec{p}_3$. Changing $x \to - x$, one obtains the $u$-channel amplitude:
\begin{align}
\label{u-channel-acc}
	iM_u =  \frac{i \Big\{E^3(x^2-1) + 8Ea^2H^2\big(3-2x-4x^2- \frac{2H^2 M_{Pl}^2(x+2)}{\Sigma}\big)\Big\}^2}{1024 \epsilon M_{Pl}^2a^6H^4(1+x)},
\end{align} 
and the diagram for this process is the right one in Fig. \ref{fig:channels}.

The matrix elements for $t$- and $u$-channels can be obtained straightforwardly (though the calculations are quite cumbersome; we present some clarifications about the corresponding calculations in Appendix~\ref{app:t_u_ch}), while the $s$-channel element calculation involves some subtlety, which is related to the terms with $\psi$ factor \eqref{psi} in the cubic order action \eqref{cubic}. We discuss this subtlety and clarify how to deal with it in Appendix~\ref{app:s_ch}. 
Before turning to the partial wave amplitude, we note that $M_t$ and $M_u$ are suppressed by $\epsilon$ as compared to $M_s$, so we will use $M \approx M_s$, where initially $M$ is the full matrix element, given by the sum of all channels amplitudes.
Finally, we find the PWA \eqref{PWA} with $l = 0$, which provides the lowest bound on the amplitude
\begin{align}
\label{a0_accur}
    \tilde{a}_0 = -E^2\sqrt{\Sigma}\cdot\frac{\big(E^4\Sigma^2 + E^2 a^2H^2\Sigma(-8 H^2 M_{Pl}^2+5\Sigma) + 4 a^4 H^4(-2H^2 M_{Pl}^2+\Sigma)^2\big)}{8192\pi\epsilon^{7/2}M_{Pl}^7a^6H^9},
\end{align}
and corresponding strong coupling energy scale can be found from the unitarity bound
\begin{align}
\label{unit_bound_0}
    |\tilde{a}_0| = \frac{1}{2}.
\end{align}
In Section \ref{sec:concrete_model} we will choose a specific model of k-inflation and obtain the concrete constraint on model parameters. If the parameters of the model satisfy these constraint then the classical description is valid.

\section{Primordial non-Gaussianities}
\label{sec:nongauss}
Another conditions on the parameters of the model of k-inflation with the lagrangian \eqref{Hor_L} comes from the observational constraints on  primordial scalar non-Gaussianities. 
The extent of non-Gaussianity can be quantified by evaluating the bispectrum of curvature perturbations $\mathcal{R}$, as
\begin{align*}
&\left\langle\mathcal{R}(\vec{k}_{1}) \mathcal{R}(\vec{k}_{2}) \mathcal{R}(\vec{k}_{3})\right\rangle\nonumber\\
&=(2 \pi)^{3} \delta^{(3)}\left(\vec{k}_{1}+\vec{k}_{2}+\vec{k}_{3}\right) B_{\mathcal{R}}\left(k_{1}, k_{2}, k_{3}\right),
\end{align*}
where $\mathcal{R}(\vec{k})$ is a Fourier component of $\mathcal{R}$ with a wave number $\vec{k}$ and the bispectrum is
\begin{align*}
    B_{\mathcal{R}}\left(k_{1}, k_{2}, k_{3}\right)=\frac{(2 \pi)^{4}\left(\mathcal{P}_{\mathcal{R}}\right)^{2}}{\prod_{i=1}^{3} k_{i}^{3}} \mathcal{A}_{\mathcal{R}}\left(k_{1}, k_{2}, k_{3}\right),
\end{align*}
which translates into a non-linear parameter $f_{\text{NL}}$ as
\begin{equation*}
    f_{NL} = \frac{10}{3}\frac{\mathcal{A}_{\mathcal{R}}}{\sum^{3}_{i=1}k_i^3},
\end{equation*}
where $\mathcal{P}_{\mathcal{R}}$ is a power spectrum and $\mathcal{A}_{\mathcal{R}}$ being its amplitude.
The bispectrum can be of different forms depending on the relation between the $\vec{k}_1, \vec{k}_2, \vec{k}_3$. In this paper, we stick to the well-known equilateral configuration $f_{\text{NL}}^{\text{equil}}$ with $k_1 = k_2 =k_3$.
The corresponding calculations of the scalar non-Gaussianities for the k-inflation with the lagrangian \eqref{Hor_L} are given in Ref. \cite{DeFelice:2011zh}. The non-linear parameter $f^{\text{equil}}_{\text{NL}}$  for the equilateral form is given by \cite{DeFelice:2011zh} 
\begin{align}
\label{fnl_felice}
    &f_{\text{NL}}^{\text{equil}} = \frac{85}{324} \Big(1-\frac{1}{c_S^2}\Big) - \frac{10}{81}\frac{\lambda}{\Sigma} \nonumber\\
    &+ \frac{55}{36} \frac{\epsilon}{c_S^2} + \frac{5}{12}\frac{\eta}{c_S^2}-\frac{85}{54}\frac{s}{c_S^2},
\end{align}
where the following notations were used:
\begin{align*}
    &\eta \equiv \dot{\epsilon}/(H\epsilon),\\
    &s \equiv \dot{c}_S/(Hc_S),\\
    &\lambda \equiv X^2 G_{2XX} + 2 X^3 G_{2XXX}/3,
\end{align*} 
and $\eta\ll 1$, $s\ll 1$, while $\lambda$ is generally not small.  In the next Section we choose a concrete model of k-inflation and find the specific form of conditions of model parameters coming from primordial scalar non-Gaussianities.

\section{Constraints on the model parameters from strong coupling problem and scalar non-Gaussianities}
\label{sec:concrete_model}
In this Section we choose a concrete model of k-inflation and show that some non-trivial condition on the parameter of the model indeed arises from the requirement of the classical description validity. 
To this end, we take the lagrangian \eqref{Hor_L} with    
\begin{equation}
\label{specific_G2}
    G_2(\phi,X) =  - \frac{16M_{Pl}^2}{9\gamma^2\phi^2}X + \frac{16 M_{Pl}^2}{9\gamma^2\phi^2M^2} X^2,  
\end{equation}
where $\gamma$ is a parameter with $[\gamma] =0$, and $M$ is another dimensional parameter, $[M]=1$; the setup with eq.~\eqref{specific_G2} is similar to the one from Ref.~\cite{Chen:2006nt}. 
For this model equations of motion \eqref{EoM} provide 
\begin{align}
\label{H_X_sol}
    H = \frac{2M}{3\sqrt{3}\gamma\phi}, \quad
    X = \frac{M^2}{2},
\end{align}
and for the scalar field we obtain
\begin{equation}
\label{phi_sol}
    \phi = Mt+c,
\end{equation}
choosing $\phi>0$ during $0<t<+\infty$, without the loss of generality. 
Here $c$ is a dimensionless constant.
After that we find all other functions and they read:
\begin{equation*}
    \Sigma = \frac{16M^2M_{Pl}^2}{9\gamma^2\phi^2},
\end{equation*}
\begin{align}
\label{csg}
    \mathcal{G}_S = 12 M^{2}_{Pl},\quad c_S^2 = \frac{\sqrt{3}}{8}\gamma,
\end{align}
so $\gamma >0$ due to the stability requirement \eqref{stability_conditions}.
Also, the slow roll parameter \eqref{eps} for the model \eqref{specific_G2} is
\begin{equation}
\label{concrete_eps}
    \epsilon = \frac{3\cdot \sqrt{3}\cdot \gamma}{2}\ll 1,
\end{equation}
which provides that $\gamma \ll 1$ as well as $c_S^2\ll 1$. This situation is similar to Ref.~\cite{Ageeva:2022asq}, so this justifies that scalar sector provides the strongest conditions of classical description validity.  

In the considered model of k-inflation the cosmological perturbations with a slightly red-tilted power spectrum may be generated \cite{Chen:2006nt}. The power spectrum of $\mathcal{R}$ perturbations is given by \cite{DeFelice:2011zh}:
\begin{align}
\label{power_sp}
    \mathcal{P}_{\mathcal{R}} = \mathcal{A}_{\mathcal{R}}\Big(\frac{k}{k_{*}}\Big)^{n_S-1}=\frac{H^2}{8\pi^2\mathcal{G}_Sc_S^3} ,
\end{align}
where $\mathcal{A}_{\mathcal{R}}$ is an amplitude, $k_*$ is a pivot momentum, $n_S$ is a spectral tilt. Surely, we require that the exit beyond effective horizon must occur in the weak coupling regime. To this end we turn to unitarity bound 
 to see whether this condition can be satisfied at the times when the relevant modes of perturbations exit the effective horizon. The corresponding PWA \eqref{a0_accur} in the chosen model \eqref{specific_G2} reads
\begin{align}
\label{a0_spec}
    \tilde{a}_0 = -\frac{\mathcal{E}^2\big(400 + 4212 \cdot\gamma^2(\frac{\mathcal{E}}{M})^2 (M t+c)^2 + 6561 \cdot\gamma^4 (\frac{\mathcal{E}}{M})^4(M t+c)^4 \big)}{20736 \cdot \sqrt{2} \cdot 3^{3/4} \cdot \pi\cdot \gamma^{7/2} M_{Pl}^2},
\end{align}
where we substitute eqs. \eqref{H_X_sol}, \eqref{phi_sol}, and \eqref{csg} and we also turn to $\mathcal{E} = E/a$, i.e. the energy in cosmic time $t$. Introducing the change of variables $\tilde{\mathcal{E}} \equiv \gamma^2(\frac{\mathcal{E}}{M})^2  (M t+c)^2$, one can sufficiently simplify \eqref{a0_spec}:
\begin{align}
\label{a0_spec_simple}
    \tilde{a}_0 = -\frac{\tilde{\mathcal{E}}M^2\big(400 + 4212 \cdot \tilde{\mathcal{E}} + 6561 \cdot\tilde{\mathcal{E}}^2\big)}{20736 \cdot \sqrt{2} \cdot 3^{3/4} \cdot \pi\cdot \gamma^{11/2} M_{Pl}^2 (M t+c)^2}.
\end{align}
  To obtain a rough estimate, we find the exit time $t_f$ at $k = k_*$, keeping in mind the smallness of $|n_S-1|$:
\begin{equation*}
    (M t_{f} + c)^2 = \frac{h_0^2}{8\pi^2 \mathcal{G}_Sc_S^3\mathcal{A}_{\mathcal{R}}},\quad h_0 = \frac{2M}{3\sqrt{3}\gamma}, 
\end{equation*}
where we use eqs.~\eqref{H_X_sol}, \eqref{phi_sol}, \eqref{csg}, and \eqref{power_sp}.

At $t_f$ eq.~\eqref{a0_spec_simple} takes quite simple form
\begin{equation*}
    \tilde{a}_0 = -\frac{ \pi\mathcal{A}_{\mathcal{R}}\tilde{\mathcal{E}}}{1024 \gamma^2}\big(400 + 4212 \cdot \tilde{\mathcal{E}} + 6561 \cdot\tilde{\mathcal{E}}^2\big),
\end{equation*}
where the observational value of $\mathcal{A}_{\mathcal{R}}=2\cdot 10^{-9}$ \cite{Planck:2019kim}.
The boundary value of  $\tilde{\mathcal{E}}$ at the classical scale $\mathcal{E} = H$ is given by
\begin{align*}
    \tilde{\mathcal{E}}\Big|_{H} = \gamma^2\Big(\frac{H}{M}\Big)^2  (M t+c)^2 = \frac{4}{27},
\end{align*}
where we use eqs.~\eqref{H_X_sol} and  \eqref{phi_sol}. The classical description is valid when $\tilde{\mathcal{E}} \gg \frac{4}{27}$ as well as $|\tilde{a}_0| \leq \frac{1}{2}$, so these two conditions lead to the constraint for $\gamma$ 
\begin{equation}
\label{sc_final}
    \gamma \gg 4.6\cdot 10^{-5}.
\end{equation}  

Next, one can obtain an additional condition on the k-inflation model parameter based on the current experimental bounds for scalar non-Gaussianities, i.e. $f_{\text{NL}}^{\text{equil}} = -26\pm 47$ \cite{Planck:2019kim}.
The leading term from eq.~\eqref{fnl_felice} is
\begin{align*}
    f_{\text{NL}}^{\text{equil}} \approx -\frac{85}{324c_S^2},
\end{align*}
and in the model with \eqref{specific_G2} and $c_S^2$ given by eq.~\eqref{csg} we obtain
\begin{equation*}
    f_{\text{NL}}^{\text{equil}} \approx -\frac{1.2}{\gamma},
\end{equation*}
where the behaviour $\sim 1/\gamma$ coincides with Ref.~\cite{Chen:2006nt}. The observational value of $f_{\text{NL}}^{\text{equil}}$ has an error larger than the value itself, i.e. $f_{\text{NL}}^{\text{equil}} = -26\pm 47$, for $68$ \% CL  \cite{Planck:2019kim}. Thus, let us choose the biggest confidence region, for example, $99.7$ \% CL from Ref.~\cite{Planck:2019kim} (see Fig. 19 therein). Roughly, this confidence region provides the constraint $|f_{\text{NL}}^{\text{equil}}|<180$, so:
\begin{equation}
\label{ng_final}
    \gamma > 0.0067.
\end{equation}
The result is as follows: in the considered model of k-inflation the absence of strong coupling problem \eqref{sc_final} is guaranteed in the presence of the observational bound for scalar non-Gaussianities \eqref{ng_final}. 

Finally, let us find other constraints coming from the calculations of the spectral tilt $n_S$ and $r$-ratio. We start with the spectral tilt, which reads \cite{DeFelice:2011zh}
\begin{equation*}
    n_S - 1 \approx -2 \epsilon = -3\sqrt{3}\gamma,
\end{equation*}
where for the second equality we have substituted $\epsilon$ from eq.~\eqref{concrete_eps}.
For the observational values $n_S = 0.9649\pm 0.0042$ \cite{Planck:2018vyg} the corresponding $\gamma$ satisfies 
\begin{equation}
\label{gamm_ns}
    0.0060<\gamma < 0.0075.
\end{equation}
Next and final constraint comes from the observational upper bound on $r$-ratio \cite{DeFelice:2011zh,Planck:2018vyg,BICEP:2021xfz}:
\begin{equation}
\label{r}
    r \equiv \frac{\mathcal{P}_T}{\mathcal{P}_{\mathcal{R}}} = \frac{4 \mathcal{G}_Sc_S^3}{\mathcal{G}_Tc_T^3} ,
\end{equation}
where $\mathcal{G}_T$ is a coupling from second order action for tensor perturbations
\begin{equation*}
    \mathcal{S}_{T} = \sum_{\lambda} \int dt d^3x a^3 \mathcal{G}_T \Big[\dot{h}_{\lambda} - \frac{c_T^2}{a^2}(\partial h_{\lambda})^2\Big], 
\end{equation*}
and $c_T$ is a tensor perturbation sound speed; $\lambda$ means two polarization of tensor perturbation. The corresponding  experimental upper bound is \cite{Planck:2018vyg,BICEP:2021xfz,Tristram:2021tvh}
\begin{equation}
\label{r_exp}
    r < 0.032.
\end{equation}
For the model \eqref{specific_G2} we have
\begin{equation*}
    \mathcal{G}_T = \frac{1}{4} M_{Pl}^2,
\end{equation*}
and so $r$-ratio \eqref{r} is
\begin{equation*}
    r = 6 \cdot 3^{3/4} \cdot \sqrt{2}\cdot  \gamma^{3/2},
\end{equation*}
where we also substitute eq.~\eqref{csg}. Finally, applying eq.~\eqref{r_exp} we arrive to
\begin{equation}
\label{gamma_r}
    \gamma < 0.014.
\end{equation}

Thus we conclude, that the strongest conditions coming both from the observational bounds on  $n_S$ and $f_{\text{NL}}^{\text{equil}}$ are
\begin{equation*}
0.0067<\gamma<0.0075.
\end{equation*}
However, if one takes another confidence region when calculating $f_{\text{NL}}^{\text{equil}}$, for example $68$ \% CL, see Ref.~\cite{Planck:2019kim}, then the model of k-inflation with \eqref{specific_G2} will be ruled out due to the inconsistency among the conditions from  $n_S$, $r$-ratio and non-Gaussianities.

\section{Conclusion}
\label{sec:conclusion}

This paper demonstrates that the specific model of k-inflation \eqref{Hor_L} with $G_2$ given by eq. \eqref{specific_G2} meets strong coupling problem. However, it is possible to find such parameters of the model that the approach of classic field theory is legitimate during considered k-inflation. We prove this statement with the proceeding to the accurate analysis of $2\to2$ processes and corresponding matrix elements, and then apply unitarity bound in order to obtain a non-trivial condition of the model parameter $\gamma$. Another constraint comes from the recent observational data for scalar non-Gaussianities from Planck \cite{Planck:2019kim}. We find out that the latter is much stronger than the condition from strong coupling absence. Thus, we conclude that the model of k-inflation with \eqref{specific_G2} is healthy: choosing $\gamma$ parameter from the permitted area, one obtains a stable theory, where the classical description is valid, and corresponding $f_{\text{NL}}^{\text{equil}}$ is allowed by the current experimental bounds. The calculations of scattering amplitudes with the use of conservation laws for energy and momentum are reasonable, since the action for scalar has the canonical form \eqref{second_canonic} at the considered times. The latter statement is proved in Appendix \ref{app:second_order}. We should also note, that for  simplicity we make our calculations for the scalar sector of primordial perturbations only, however, there are mixed and tensor sectors as well. We expect that, as usual (see, for instance Ref. \cite{Ageeva:2022asq}), these sectors give even weaker constraints than the scalar one. There remains another important question: does the same result hold for each known (and phenomenologically interesting) model of inflation or maybe it is not the case? For example,  models of $G$-inflation contain higher order partial derivatives in the cubic action for scalars, thus, it potentially can strengthen strong coupling absence condition.

\section*{Acknowledgments} 

The authors are thankful to Valery Rubakov for valuable discussions and very useful comments at early stages of this work back in the days. The authors are grateful to Mikhail Shaposhnikov, Eugeny Babichev, Pavel Demidov, Sergei Demidov, Sergei Mironov, Petr Satunin, Vladislav Barinov, Andrei Kataev and Victoria Volkova
for fruitful discussions and careful reading of this manuscript. This work has been 
supported by Russian Science Foundation Grant No. 19-12-00393.

\appendix

\section{The canonical second order action for scalar perturbation}
\label{app:second_order}
\numberwithin{equation}{section}
In this Appendix we comment why we indeed can use second order action in the canonical form  \eqref{second_canonic} at the early times (where we study the strong coupling issue) and thus consider the energy conservation laws (which we require to calculate the scattering amplitudes) for the mode or for the pair of modes. 

We begin with the second order action for $\mathcal{R}$, which is given by \eqref{second}. 
However, it is more clear to work in terms of conformal time $dt = a d\eta$, thus
\begin{equation}
\label{2action_tR}
    \mathcal{S}^{(2)}_{\mathcal{R}\mathcal{R}} = \int a^2 d^3x d\eta \mathcal{G}_S \left(\mathcal{R}^{\prime\; 2} - c_S^2(\vec{\nabla} \mathcal{R})^2\right).
\end{equation}
Another convenient change of variables is $\tilde{\mathcal{R}} = z \mathcal{R}$ with $z = a\sqrt{2\mathcal{G}_S}$, so the action \eqref{2action_tR} has the form
\begin{align*}
    \mathcal{S}^{(2)}_{\tilde{\mathcal{R}}\tilde{\mathcal{R}}} &= \frac{1}{2}\int d^3x d\eta \Big(\tilde{\mathcal{R}}^{\prime\; 2} - \frac{2 z^{\prime} \tilde{\mathcal{R}} \tilde{\mathcal{R}}^{\prime}}{z} \nonumber\\
    &+ \frac{z^{\prime 2}\tilde{\mathcal{R}}^{2}}{z^2} - c_S^2(\vec{\nabla} \tilde{\mathcal{R}})^2\Big).
\end{align*}
After some integration by parts for the second term in the latter formula we arrive to
\begin{equation}
    \label{s2_intermed}
\mathcal{S}^{(2)}_{\tilde{\mathcal{R}}\tilde{\mathcal{R}}} = \frac{1}{2}\int d^3x d\eta \left(\tilde{\mathcal{R}}^{\prime\; 2} + \frac{z^{\prime \prime}\tilde{\mathcal{R}}^{2}}{z} - c_S^2(\vec{\nabla} \tilde{\mathcal{R}})^2\right).
\end{equation}
The coefficient in the second term is  $\frac{z''}{z} \approx 2H^2a^2$, where we neglect with $a H'$ contribution due to $\epsilon \ll 1$ (and we also suppose that $\mathcal{G}_S$ is slowly varying with time). Indeed, using $\epsilon = -H'/aH^2$ one can obtain
\begin{align}
    \frac{z''}{z} = 2  a' H + a H' = 2 a^2 H^2 - \epsilon a^2 H^2 ,
\end{align}
where the last term surely can be omitted.
 We also assume that we work at such energies that much bigger than classical energy scale (i.e. Hubble parameter), thus we can omit the second term in \eqref{s2_intermed} and finally arrive to the canonical form \eqref{second_canonic}.

\section{Expressions for $\Lambda_{i,(j)}$}
\label{app:lambdas}
\numberwithin{equation}{section}

\setcounter{equation}{0}
The purpose of this Appendix is to list the cubic order action coefficients from eq.~\eqref{cubic}. The general expressions for $\Lambda_{i}$, with $i = 1,\ldots,17$ are given in Ref. \cite{Ageeva:2020gti}, and the formulas for the specific model \eqref{Hor_L} read:
\renewcommand{\arraystretch}{2.05}
\[
   \begin{array}{*2{>{\displaystyle}r>{\displaystyle}l}}
        \Lambda_1 [\dot{\mathcal{R}}^3] 
        &\multicolumn{2}{l}{\displaystyle = \frac{3\Sigma^2-2M^2_{Pl}H^2X\big(3G_{2X}+4X(3G_{2XX}+XG_{2XXX})\big)}{6M^2_{Pl}H^5}, \nonumber } & \\
        \Lambda_2 [\dot{\mathcal{R}}^2\mathcal{R}] &\multicolumn{2}{l}{\displaystyle = -\frac{3\Sigma\big(-2M^2_{Pl}H^2+\Sigma\big)}{2M^2_{Pl}H^4},  \nonumber } & \\
        \Lambda_3 [(\dot{\mathcal{R}}^2/a^2) \partial^2 \mathcal{R}] &\multicolumn{2}{l}{\displaystyle = -\frac{\Sigma}{H^4}, \nonumber } & \\
        \Lambda_4 [(\dot{\mathcal{R}}/a^2)\mathcal{R} \partial^2 \mathcal{R}] &=\frac{-2M^2_{Pl}H^2+3\Sigma}{H^3}, \\
        \Lambda_5 [(\dot{\mathcal{R}}/a^2) \left(\partial_i \mathcal{R} \right)^2] &=
        \frac{-M^2_{Pl}H^2+2\Sigma}{H^3}, \nonumber
        \\
        \Lambda_6 [(\mathcal{R}/a^2) \left(\partial_i \mathcal{R} \right)^2] &=M^2_{Pl},  
        & \Lambda_7 [(\dot{\mathcal{R}}/a^4) \left(\partial^2 \mathcal{R} \right)^2] 
        &=\frac{M^2_{Pl}}{2H^3}, \nonumber \\
        \Lambda_8[(\mathcal{R}/a^4) \left(\partial^2 \mathcal{R} \right)^2]&=
        -\frac{3M^2_{Pl}}{2H^2}, 
        & \Lambda_9[(\partial^2 \mathcal{R}/a^4) \left(\partial_i \mathcal{R} \right)^2]&=
        -\frac{2M^2_{Pl}}{H^2},  \nonumber \\
        \Lambda_{10}[(\dot{\mathcal{R}}/a^4) \big( \partial_i \partial_j \mathcal{R} \big)^2]&=
        -\frac{M^2_{Pl}}{2H^3},  
        & \Lambda_{11}[(\mathcal{R}/a^4) \big( \partial_i \partial_j \mathcal{R} \big)^2]&=\frac{3M^2_{Pl}}{2H^2},   \nonumber \\
        \Lambda_{12}[\dot{\mathcal{R}} \partial_i \mathcal{R} \partial^i \psi]&=
        -\frac{2\Sigma^2}{M^2_{Pl}H^4}, 
        & \Lambda_{13}[(\partial^2 \mathcal{R}/a^2) \partial_i \mathcal{R} \partial^i \psi]&=
        \frac{2\Sigma}{H^3},  \nonumber \\
        \Lambda_{14}[\dot{\mathcal{R}} \big( \partial_i \partial_j \psi \big)^2]&=
        -\frac{\Sigma^2}{2M^2_{Pl}H^5},  
        & \Lambda_{15}[\mathcal{R} \big( \partial_i \partial_j \psi \big)^2]&=
        \frac{3\Sigma^2}{2M^2_{Pl}H^4}, \nonumber \\
        \Lambda_{16}[(\dot{\mathcal{R}}/a^2) \partial_i \partial_j \mathcal{R} \partial^i \partial^j \psi]&=
        \frac{\Sigma}{H^4},  
        & \Lambda_{17}[(\mathcal{R}/a^2) \partial_i \partial_j \mathcal{R} \partial^i \partial^j \psi]&=
        -\frac{3\Sigma}{H^3}.
	\end{array}
\]
One can rewrite the expressions above, using eq.~\eqref{cs} and introducing
\begin{align*}
    \lambda_1 \equiv X^2 G_{2XX} + X^3 G_{2XXX}/3,
\end{align*}
thus, we arrive to
\renewcommand{\arraystretch}{2.05}
\[
   \begin{array}{*2{>{\displaystyle}r>{\displaystyle}l}}
        \Lambda_1 [\dot{\mathcal{R}}^3] 
        &\multicolumn{2}{l}{\displaystyle = \frac{\Big(3M^2_{Pl}\big(\frac{H^2\epsilon}{c_S^2}\big)^2-2H^2\big[3 \epsilon M^2_{Pl} H^2+12\lambda_1\big]\Big)}{6H^5}, \nonumber } & \\
        \Lambda_2 [\dot{\mathcal{R}}^2\mathcal{R}] &\multicolumn{2}{l}{\displaystyle = -\frac{3M_{Pl}^2\epsilon}{2c_S^2}\Big(-2+\frac{\epsilon}{c_S^2}\Big),  \nonumber } & \\
        \Lambda_3 [(\dot{\mathcal{R}}^2/a^2) \partial^2 \mathcal{R}] &\multicolumn{2}{l}{\displaystyle = -\frac{M_{Pl}^2\epsilon}{c_S^2H^2}, \nonumber } & \\
        \Lambda_4 [(\dot{\mathcal{R}}/a^2)\mathcal{R} \partial^2 \mathcal{R}] &=\frac{-2M_{Pl}^2+\frac{3M_{Pl}^2\epsilon}{c_S^2}}{H}, \\
        \Lambda_5 [(\dot{\mathcal{R}}/a^2) \left(\partial_i \mathcal{R} \right)^2] &=
        \frac{-M_{Pl}^2+\frac{2M_{Pl}^2\epsilon}{c_S^2}}{H}, \nonumber
        \\
        \Lambda_6 [(\mathcal{R}/a^2) \left(\partial_i \mathcal{R} \right)^2] &=M_{Pl}^2,  
        & \Lambda_7 [(\dot{\mathcal{R}}/a^4) \left(\partial^2 \mathcal{R} \right)^2] 
        &=\frac{M_{Pl}^2}{2H^3}, \nonumber \\
       \Lambda_8[(\mathcal{R}/a^4) \left(\partial^2 \mathcal{R} \right)^2]&=
        -\frac{3M_{Pl}^2}{2H^2}, 
        &\Lambda_9[(\partial^2 \mathcal{R}/a^4) \left(\partial_i \mathcal{R} \right)^2]&=
        -\frac{2M_{Pl}^2}{H^2},  \nonumber \\
        \Lambda_{10}[(\dot{\mathcal{R}}/a^4) \big( \partial_i \partial_j \mathcal{R} \big)^2]&=
        -\frac{M_{Pl}^2}{2H^3},  
        & \Lambda_{11}[(\mathcal{R}/a^4) \big( \partial_i \partial_j \mathcal{R} \big)^2]&=\frac{3M_{Pl}^2}{2H^2},   \nonumber \\
        \Lambda_{12}[\dot{\mathcal{R}} \partial_i \mathcal{R} \partial^i \psi]&=
        -2M_{Pl}^2\Big(\frac{\epsilon}{c_S^2}\Big)^2, 
        & \Lambda_{13}[(\partial^2 \mathcal{R}/a^2) \partial_i \mathcal{R} \partial^i \psi]&=
        \frac{2M_{Pl}^2\epsilon}{c_S^2H},  \nonumber \\
        \Lambda_{14}[\dot{\mathcal{R}} \big( \partial_i \partial_j \psi \big)^2]&=
        -\frac{M_{Pl}^2(\frac{\epsilon}{c_S^2})^2}{2H},  
        & \Lambda_{15}[\mathcal{R} \big( \partial_i \partial_j \psi \big)^2]&=
        \frac{3M_{Pl}^2(\frac{\epsilon}{c_S^2})^2}{2}, \nonumber \\
         \Lambda_{16}[(\dot{\mathcal{R}}/a^2) \partial_i \partial_j \mathcal{R} \partial^i \partial^j \psi]&=
        \frac{M_{Pl}^2(\frac{\epsilon}{c_S^2})}{H^2},  
        & \Lambda_{17}[(\mathcal{R}/a^2) \partial_i \partial_j \mathcal{R} \partial^i \partial^j \psi]&=
        -\frac{3M_{Pl}^2(\frac{\epsilon}{c_S^2})}{H}.  
	\end{array}
\] 
Using these expressions, as well as keeping in mind the discussion about $\Lambda_{7}-\Lambda_{11}$, see eqs. \eqref{l7-l11}-\eqref{L_7891011}, we substitute field $\mathcal{R} = \tilde{\mathcal{R}}/z$ into eq. \eqref{cubic} and obtain:
\begin{align}
\label{app:cubic_canonic}
        \mathcal{S}^{(3)}_{\mathcal{R}\mathcal{R}\mathcal{R}}&=   \int d\eta\text{ }d^3x \Big\{\Lambda_{1(1)}\tilde{\mathcal{R}}^{\prime\;3}  + \Lambda_{1(2)}\tilde{\mathcal{R}}\tilde{\mathcal{R}}^{\prime\;2}+\Lambda_{1(3)}\tilde{\mathcal{R}}^2\tilde{\mathcal{R}}^{\prime}+\Lambda_{1(4)}\tilde{\mathcal{R}}^3   \nonumber\\ 
        &+\Lambda_{2(1)} \tilde{\mathcal{R}}\tilde{\mathcal{R}}^{\prime\;2}  +\Lambda_{2(2)} \tilde{\mathcal{R}}^2\tilde{\mathcal{R}}^{\prime} + \Lambda_{2(3)}\tilde{\mathcal{R}}^3  \nonumber\\  
        &+\Lambda_{3(1)} \tilde{\mathcal{R}}^2 \partial^2 \tilde{\mathcal{R}} +\Lambda_{3(2)} \tilde{\mathcal{R}}\tilde{\mathcal{R}}^{\prime}\partial^2 \tilde{\mathcal{R}} + \Lambda_{3(3)}\tilde{\mathcal{R}}^{\prime\;2}\partial^2 \tilde{\mathcal{R}}  \nonumber\\  
        &+\Lambda_{4(1)} \tilde{\mathcal{R}}^2  \partial^2 \tilde{\mathcal{R}} + \Lambda_{4(2)} \tilde{\mathcal{R}}\tilde{\mathcal{R}}^{\prime}\partial^2 \tilde{\mathcal{R}}\nonumber\\  &+\Lambda_{5(1)}\tilde{\mathcal{R}}(\partial_i \tilde{\mathcal{R}})^2 + \Lambda_{5(2)}\tilde{\mathcal{R}}^{\prime}(\partial_i \tilde{\mathcal{R}})^2\nonumber\\
        &+\Lambda_{6(1)} \tilde{\mathcal{R}}(\partial_i \tilde{\mathcal{R}})^2  
        \nonumber\\
        &+ \Lambda_{*(1)}\tilde{\mathcal{R}} \left( (\partial^2 \tilde{\mathcal{R}} )^2-(\partial_i \partial_j \tilde{\mathcal{R}} )^2\right)  
        \nonumber\\
        &+ 
\Lambda_{12(1)}\partial_i\tilde{\mathcal{R}}\tilde{\mathcal{R}}^{\prime} \partial_i \partial^{-2} \tilde{\mathcal{R}}^{\prime}+ \Lambda_{12(2)}\tilde{\mathcal{R}}^{\prime}\partial_i\tilde{\mathcal{R}} \partial_i \partial^{-2} \tilde{\mathcal{R}}+ \Lambda_{12(3)}\tilde{\mathcal{R}}\partial_i\tilde{\mathcal{R}} \partial_i \partial^{-2} \tilde{\mathcal{R}}^{\prime}+ \Lambda_{12(4)}\tilde{\mathcal{R}} \partial_i\tilde{\mathcal{R}} \partial_i \partial^{-2} \tilde{\mathcal{R}}\nonumber\\
        &+ \Lambda_{13(1)} \partial^2 \tilde{\mathcal{R}}\partial_i \tilde{\mathcal{R}}\partial_i \partial^{-2} \tilde{\mathcal{R}} + \Lambda_{13(2)}\partial^2 \tilde{\mathcal{R}}\partial_i \tilde{\mathcal{R}}\partial_i \tilde{\psi} \nonumber\\
        &+ 
      \Lambda_{14(1)}\tilde{\mathcal{R}}^{\prime} (\partial_i\partial_j \partial^{-2} \tilde{\mathcal{R}}^{\prime})^2+ \Lambda_{14(2)}\tilde{\mathcal{R}}^{\prime}\partial_i\partial_j\partial^{-2}\tilde{\mathcal{R}} \partial_i\partial_j \partial^{-2}  \tilde{\mathcal{R}}^{\prime}\nonumber\\
      &+ \Lambda_{14(3)}\tilde{\mathcal{R}}(\partial_i\partial_j\partial^{-2}\tilde{\mathcal{R}}^{\prime})^2 + \Lambda_{14(4)}\tilde{\mathcal{R}}^{\prime}(\partial_i\partial_j\partial^{-2}\tilde{\mathcal{R}})^2 \nonumber\\
      &+ \Lambda_{14(5)}\tilde{\mathcal{R}}\partial_i\partial_j\partial^{-2}\tilde{\mathcal{R}}\partial_i\partial_j\partial^{-2}\tilde{\mathcal{R}}^{\prime} + \Lambda_{14(6)}\tilde{\mathcal{R}}(\partial_i\partial_j\partial^{-2}\tilde{\mathcal{R}})^2
      \nonumber\\
      &+ 
\Lambda_{15(1)}\tilde{\mathcal{R}}(\partial_i\partial_j\partial^{-2}\tilde{\mathcal{R}}^{\prime})^2+ \Lambda_{15(2)}\tilde{\mathcal{R}}\partial_i\partial_j\partial^{-2}\tilde{\mathcal{R}} \partial_i\partial_j  \partial^{-2}\tilde{\mathcal{R}}^{\prime}+ \Lambda_{15(3)}\tilde{\mathcal{R}}(\partial_i\partial_j\partial^{-2}\tilde{\mathcal{R}})^2\nonumber\\
        &+ 
      \Lambda_{16(1)}\tilde{\mathcal{R}}\partial_i\partial_j\tilde{\mathcal{R}} \partial_i\partial_j \partial^{-2} \tilde{\mathcal{R}}+ \Lambda_{16(2)}\tilde{\mathcal{R}}\partial_i\partial_j\tilde{\mathcal{R}} \partial_i\partial_j  \tilde{\psi}\nonumber\\
      &+ \Lambda_{16(3)}\tilde{\mathcal{R}}^{\prime}\partial_i\partial_j\tilde{\mathcal{R}} \partial_i\partial_j \partial^{-2} \tilde{\mathcal{R}}+ \Lambda_{16(4)}\tilde{\mathcal{R}}^{\prime}\partial_i\partial_j\tilde{\mathcal{R}} \partial_i\partial_j  \tilde{\psi} \nonumber\\
        &+ \Lambda_{17(1)} \tilde{\mathcal{R}} \partial_i\partial_j\tilde{\mathcal{R}} \partial_i\partial_j \partial^{-2} \tilde{\mathcal{R}}+ \Lambda_{17(2)} \tilde{\mathcal{R}} \partial_i\partial_j\tilde{\mathcal{R}} \partial_i\partial_j \tilde{\psi}\Big\},
    \end{align}
where
\begin{equation*}
    \Lambda_{1(1)} =  \frac{\Lambda_{1}}{2\sqrt{2}\mathcal{G}_S^{3/2}a^2},\quad \Lambda_{1(2)} = - \frac{3\Lambda_{1}H}{2\sqrt{2}\mathcal{G}_S^{3/2}a}, \quad \Lambda_{1(3)} =  \frac{3\Lambda_{1}H^2}{2\sqrt{2}\mathcal{G}_S^{3/2}}, \quad \Lambda_{1(4)} =  -\frac{\Lambda_{1}aH^3}{2\sqrt{2}\mathcal{G}_S^{3/2}},
\end{equation*}
\begin{equation*}
    \Lambda_{2(1)} =  \frac{\Lambda_{2}}{2\sqrt{2}\mathcal{G}_S^{3/2}a},\quad \Lambda_{2(2)} = - \frac{\Lambda_{2}H}{\sqrt{2}\mathcal{G}_S^{3/2}}, \quad \Lambda_{2(3)} =  \frac{\Lambda_{2}aH^2}{2\sqrt{2}\mathcal{G}_S^{3/2}},
\end{equation*}
\begin{equation*}
    \Lambda_{3(1)} =  \frac{\Lambda_{3}H^2}{2\sqrt{2}\mathcal{G}_S^{3/2}a},\quad \Lambda_{3(2)} = - \frac{\Lambda_{3}H}{\sqrt{2}\mathcal{G}_S^{3/2}a^2}, \quad \Lambda_{3(3)} =  \frac{\Lambda_{3}}{2\sqrt{2}\mathcal{G}_S^{3/2}a^3},
\end{equation*}
\begin{equation*}
    \Lambda_{4(1)} =  -\frac{\Lambda_{4}H}{2\sqrt{2}\mathcal{G}_S^{3/2}a},\quad \Lambda_{4(2)} =  \frac{\Lambda_{4}}{2\sqrt{2}\mathcal{G}_S^{3/2}a^2}, 
\end{equation*}
\begin{equation*}
    \Lambda_{5(1)} =  -\frac{\Lambda_{5}H}{2\sqrt{2}\mathcal{G}_S^{3/2}a},\quad \Lambda_{5(2)} =  \frac{\Lambda_{5}}{2\sqrt{2}\mathcal{G}_S^{3/2}a^2}, 
\end{equation*}
\begin{equation*}
    \Lambda_{6(1)} =  \frac{\Lambda_{6}}{2\sqrt{2}\mathcal{G}_S^{3/2}a}, 
\end{equation*}
\begin{equation*}
    \Lambda_{*(1)} =  \frac{\Lambda_{*}}{2\sqrt{2}\mathcal{G}_S^{3/2}a^3}. 
\end{equation*}
\begin{equation*}
    \Lambda_{12(1)} =  \frac{\Lambda_{12}}{2\sqrt{2}\mathcal{G}_S^{3/2}a},\quad \Lambda_{12(2)} = - \frac{\Lambda_{12}H}{2\sqrt{2}\mathcal{G}_S^{3/2}},\quad \quad \Lambda_{12(3)} = - \frac{\Lambda_{12}H}{2\sqrt{2}\mathcal{G}_S^{3/2}}, \quad \Lambda_{12(4)} =  \frac{\Lambda_{12}aH^2}{2\sqrt{2}\mathcal{G}_S^{3/2}}, 
\end{equation*}
\begin{equation*}
    \Lambda_{13(1)} =  -\frac{\Lambda_{13}H}{2\sqrt{2}\mathcal{G}_S^{3/2}a},\quad \Lambda_{13(2)} =  \frac{\Lambda_{13}}{2\sqrt{2}\mathcal{G}_S^{3/2}a^2}, 
\end{equation*}
\begin{align*}
    \Lambda_{14(1)} =  \frac{\Lambda_{14}}{2\sqrt{2}\mathcal{G}_S^{3/2}a^2},\quad &\Lambda_{14(2)} =  -\frac{\Lambda_{14}H}{\sqrt{2}\mathcal{G}_S^{3/2}a}, \quad \Lambda_{14(3)} =  -\frac{\Lambda_{14}H}{2\sqrt{2}\mathcal{G}_S^{3/2}a}, \quad \Lambda_{14(4)} =  \frac{\Lambda_{14}H^2}{2\sqrt{2}\mathcal{G}_S^{3/2}},\nonumber\\
    &\Lambda_{14(5)} =  \frac{\Lambda_{14}H^2}{\sqrt{2}\mathcal{G}_S^{3/2}}, \quad \Lambda_{14(6)} =  -\frac{\Lambda_{14}aH^3}{2\sqrt{2}\mathcal{G}_S^{3/2}}
\end{align*}
\begin{equation*}
    \Lambda_{15(1)} =  \frac{\Lambda_{15}}{2\sqrt{2}\mathcal{G}_S^{3/2}a},\quad \Lambda_{15(2)} =  -\frac{\Lambda_{15}H}{\sqrt{2}\mathcal{G}_S^{3/2}}, \quad \Lambda_{15(3)} =  \frac{\Lambda_{15}aH^2}{2\sqrt{2}\mathcal{G}_S^{3/2}}, 
\end{equation*}
\begin{equation*}
    \Lambda_{16(1)} =  \frac{\Lambda_{16}H^2}{2\sqrt{2}\mathcal{G}_S^{3/2}a},\quad \Lambda_{16(2)} = - \frac{\Lambda_{16}H}{2\sqrt{2}\mathcal{G}_S^{3/2}a^2},\quad \quad \Lambda_{16(3)} = - \frac{\Lambda_{16}H}{2\sqrt{2}\mathcal{G}_S^{3/2}a^2}, \quad \Lambda_{16(4)} =  \frac{\Lambda_{16}}{2\sqrt{2}\mathcal{G}_S^{3/2}a^3}, 
\end{equation*}
\begin{equation*}
    \Lambda_{17(1)} =  -\frac{\Lambda_{17}H}{2\sqrt{2}\mathcal{G}_S^{3/2}a},\quad \Lambda_{17(2)} =  \frac{\Lambda_{17}}{2\sqrt{2}\mathcal{G}_S^{3/2}a^2}.
\end{equation*}
We use these $\Lambda_{i,(j)}$ to naively estimate the matrix elements $M_{i,(j)}$ \eqref{M_naive}. Finally, unitarity bound and the condition of strong coupling absence for each $(\tilde{a}_0)_{i,(j)}$ \eqref{a0_naive} provide the constraints on the slow roll parameter $\epsilon$ \eqref{inequalities}.

\section{Calculations of matrix elements}
\label{app:t_u_ch}
\numberwithin{equation}{section}

\setcounter{equation}{0}

In this Appendix we show explicitly how to obtain the accurate expressions for $t$- and $u$-matrix elements from Section \ref{subsec:diagram}. The subtlety of the calculation of $s$-matrix element is shown in Appendix \ref{app:s_ch}. 

First of all, we remind that we consider only $\Lambda_3-\Lambda_{6}, \Lambda_{*}, \Lambda_{13}, \Lambda_{16}, \Lambda_{17}$ terms in further calculations, since other ones provide suppressed contribution to final matrix elements, see the discussion in Section \ref{subsec:naive}. 

Begin with the calculation of $t$-channel matrix element. Consider the upper vertex (with $\vec{p}_1$ and $\vec{p}_3$ momenta) from the center panel in Fig. \ref{fig:channels}. 

1) $\Lambda_{3(1)} \tilde{\mathcal{R}}^2 \partial^2 \tilde{\mathcal{R}}+\Lambda_{3(2)} \tilde{\mathcal{R}}\tilde{\mathcal{R}}^{\prime}\partial^2 \tilde{\mathcal{R}} + \Lambda_{3(3)}\tilde{\mathcal{R}}^{\prime\;2}\partial^2 \tilde{\mathcal{R}}$ terms give: 
\begin{align*}
    &V_3 = 2i \Lambda_{3(1)} (i \vec{p}_1)^2 +2i \Lambda_{3(1)} (-i \vec{p}_3)^2 + 2i \Lambda_{3(1)} (-i \vec{p}_h)^2\nonumber\\
    + &i \Lambda_{3(2)} (iE_3)(i \vec{p}_1)^2 + i \Lambda_{3(2)} (-iE_1)(-i \vec{p}_3)^2 + i \Lambda_{3(2)} (-iE_1)(-i \vec{p}_h)^2 + i\Lambda_{3(2)} (iE_3)(-i \vec{p}_h)^2\nonumber\\
    +&2i\Lambda_{3(3)}(-iE_1)(iE_3)(-i \vec{p}_h)^2 ,
\end{align*}
where $\vec{p}_h = \vec{p}_1-\vec{p}_3$ and we also keep in mind that $E_h = E_1 - E_3 = 0$ in this and further calculations.

2) $\Lambda_{4(1)} \tilde{\mathcal{R}}^2  \partial^2 \tilde{\mathcal{R}} + \Lambda_{4(2)} \tilde{\mathcal{R}}\tilde{\mathcal{R}}^{\prime}\partial^2 \tilde{\mathcal{R}}$ terms give: 
\begin{align*}
    &V_4=2i \Lambda_{4(1)} (-i \vec{p}_h)^2 +2i \Lambda_{4(1)} (i \vec{p}_1)^2 + 2i \Lambda_{4(1)} (-i \vec{p}_3)^2 \nonumber\\
    + &i \Lambda_{4(2)} (iE_3)(i \vec{p}_1)^2 +i \Lambda_{4(2)} (-iE_1)(-i \vec{p}_3)^2 + i \Lambda_{4(2)} (-iE_1)(-i \vec{p}_h)^2 + i \Lambda_{4(2)} (iE_3)(-i \vec{p}_h)^2 .
\end{align*}

3) $\Lambda_{5(1)}\tilde{\mathcal{R}}(\partial_i \tilde{\mathcal{R}})^2 + \Lambda_{5(2)}\tilde{\mathcal{R}}^{\prime}(\partial_i \tilde{\mathcal{R}})^2$ terms give: 
\begin{align*}
    &V_5= 2i \Lambda_{5(1)} (i \vec{p}_1, -i \vec{p}_3) +2i \Lambda_{5(1)} (i \vec{p}_1, -i \vec{p}_h)+2i \Lambda_{5(1)} (-i \vec{p}_h, -i \vec{p}_3)\nonumber\\
    &+2i \Lambda_{5(2)} (i \vec{p}_1, -i \vec{p}_h)(iE_3) +2i \Lambda_{5(2)} (-i \vec{p}_h, -i \vec{p}_3)(-iE_1).
\end{align*}

4) $\Lambda_{6(1)} \tilde{\mathcal{R}}(\partial_i \tilde{\mathcal{R}})^2$ term gives:  
\begin{align*}
    &V_6=2i \Lambda_{6(1)} (i \vec{p}_1, -i \vec{p}_3)+2i \Lambda_{6(1)} (i \vec{p}_1, -i \vec{p}_h)+2i \Lambda_{6(1)} (-i \vec{p}_h, -i \vec{p}_3)  .
\end{align*}

5) $\Lambda_{13(1)} \partial^2 \tilde{\mathcal{R}}\partial_i \tilde{\mathcal{R}}\partial_i \partial^{-2} \tilde{\mathcal{R}} + \Lambda_{13(2)}\partial^2 \tilde{\mathcal{R}}\partial_i \tilde{\mathcal{R}}\partial_i \tilde{\psi}$ terms give:  
\begin{align*}
    &V_{13} = i \Lambda_{13(1)} (i\vec{p}_1)^2 \frac{(-i \vec{p}_3, -i \vec{p}_h)}{(-i\vec{p}_h)^2} + i \Lambda_{13(1)} (-i\vec{p}_3)^2 \frac{(-i \vec{p}_h, i \vec{p}_1)}{(i\vec{p}_1)^2} + i \Lambda_{13(1)} (-i\vec{p}_h)^2 \frac{(i \vec{p}_1, -i \vec{p}_3)}{(-i\vec{p}_3)^2}\nonumber\\
    &+i \Lambda_{13(1)} (-i\vec{p}_3)^2 \frac{(i \vec{p}_1, -i \vec{p}_h)}{(-i\vec{p}_h)^2} + i \Lambda_{13(1)} (i\vec{p}_1)^2 \frac{(-i \vec{p}_h, -i \vec{p}_3)}{(-i\vec{p}_3)^2} + i \Lambda_{13(1)} (-i\vec{p}_h)^2 \frac{(-i \vec{p}_3, i \vec{p}_1)}{(i\vec{p}_1)^2}\nonumber\\
    &+i \Lambda_{13(2)} (-i\vec{p}_3)^2 \frac{(-i \vec{p}_h, i \vec{p}_1)}{(i\vec{p}_1)^2}(-iE_1) + i \Lambda_{13(2)} (-i\vec{p}_h)^2 \frac{(i \vec{p}_1, -i \vec{p}_3)}{(-i\vec{p}_3)^2}(iE_3) + i \Lambda_{13(2)} (i\vec{p}_1)^2 \frac{(-i \vec{p}_h, -i \vec{p}_3)}{(-i\vec{p}_3)^2}(iE_3) \nonumber\\
    &+ i \Lambda_{13(2)} (-i\vec{p}_h)^2 \frac{(-i \vec{p}_3, i \vec{p}_1)}{(i\vec{p}_1)^2}(-iE_1).
\end{align*}

6) $\Lambda_{16(1)}\tilde{\mathcal{R}}\partial_i\partial_j\tilde{\mathcal{R}} \partial_i\partial_j \partial^{-2} \tilde{\mathcal{R}}+ \Lambda_{16(2)}\tilde{\mathcal{R}}\partial_i\partial_j\tilde{\mathcal{R}} \partial_i\partial_j  \tilde{\psi}+ \Lambda_{16(3)}\tilde{\mathcal{R}}^{\prime}\partial_i\partial_j\tilde{\mathcal{R}} \partial_i\partial_j \partial^{-2} \tilde{\mathcal{R}}+ \Lambda_{16(4)}\tilde{\mathcal{R}}^{\prime}\partial_i\partial_j\tilde{\mathcal{R}} \partial_i\partial_j  \tilde{\psi}$ terms give:   
\begin{align*}
    &V_{16} = i \Lambda_{16(1)}  \frac{(-i \vec{p}_3, -i \vec{p}_h)^2}{(-i\vec{p}_h)^2} + i \Lambda_{16(1)}  \frac{(-i \vec{p}_h, i \vec{p}_1)^2}{(i\vec{p}_1)^2} + i \Lambda_{16(1)}  \frac{(i \vec{p}_1, -i \vec{p}_3)^2}{(-i\vec{p}_3)^2}\nonumber\\
    &+i \Lambda_{16(1)}  \frac{(i \vec{p}_1, -i \vec{p}_h)^2}{(-i\vec{p}_h)^2} + i \Lambda_{16(1)}  \frac{(-i \vec{p}_h, -i \vec{p}_3)^2}{(-i\vec{p}_3)^2} + i \Lambda_{16(1)}  \frac{(-i \vec{p}_3, i \vec{p}_1)^2}{(i\vec{p}_1)^2}\nonumber\\
    &+i \Lambda_{16(2)}  \frac{(-i \vec{p}_h, i \vec{p}_1)^2}{(i\vec{p}_1)^2}(-iE_1)+i \Lambda_{16(2)}  \frac{(-i \vec{p}_3, i \vec{p}_1)^2}{(i\vec{p}_1)^2}(-iE_1)\nonumber\\
    &+i \Lambda_{16(2)}  \frac{(-i \vec{p}_h, -i \vec{p}_3)^2}{(-i\vec{p}_3)^2}(iE_3)+i \Lambda_{16(2)}  \frac{(i \vec{p}_1, -i \vec{p}_3)^2}{(-i\vec{p}_3)^2}(iE_3)\nonumber\\
    &+i \Lambda_{16(3)} (-iE_1) \frac{(-i \vec{p}_h, -i \vec{p}_3)^2}{(-i\vec{p}_3)^2}+i \Lambda_{16(3)}(-iE_1)  \frac{(-i \vec{p}_3, -i \vec{p}_h)^2}{(-i\vec{p}_h)^2}\nonumber\\
    &+i \Lambda_{16(3)}  (iE_3)\frac{(-i \vec{p}_h, i \vec{p}_1)^2}{(i\vec{p}_1)^2}+i \Lambda_{16(3)}(iE_3)  \frac{(i \vec{p}_1, -i \vec{p}_h)^2}{(-i\vec{p}_h)^2}\nonumber\\
    &+i \Lambda_{16(4)}  (-iE_1)\frac{(-i \vec{p}_h, -i \vec{p}_3)^2}{(-i\vec{p}_3)^2}(iE_3) + i \Lambda_{16(4)}  (iE_3)\frac{(-i \vec{p}_h, i \vec{p}_1)^2}{(i\vec{p}_1)^2}(-iE_1).
\end{align*}

7) Terms $\Lambda_{17(1)} \tilde{\mathcal{R}} \partial_i\partial_j\tilde{\mathcal{R}} \partial_i\partial_j \partial^{-2} \tilde{\mathcal{R}}+ \Lambda_{17(2)} \tilde{\mathcal{R}} \partial_i\partial_j\tilde{\mathcal{R}} \partial_i\partial_j \tilde{\psi}$ give: 
\begin{align*}
     &V_{17} = i \Lambda_{17(1)}  \frac{(-i \vec{p}_3, -i \vec{p}_h)^2}{(-i\vec{p}_h)^2} + i \Lambda_{17(1)}  \frac{(-i \vec{p}_h, i \vec{p}_1)^2}{(i\vec{p}_1)^2} + i \Lambda_{17(1)}  \frac{(i \vec{p}_1, -i \vec{p}_3)^2}{(-i\vec{p}_3)^2}\nonumber\\
    &+i \Lambda_{17(1)}  \frac{(i \vec{p}_1, -i \vec{p}_h)^2}{(-i\vec{p}_h)^2} + i \Lambda_{17(1)}  \frac{(-i \vec{p}_h, -i \vec{p}_3)^2}{(-i\vec{p}_3)^2} + i \Lambda_{17(1)}  \frac{(-i \vec{p}_3, i \vec{p}_1)^2}{(i\vec{p}_1)^2}\nonumber\\
    &+i \Lambda_{17(2)}  \frac{(-i \vec{p}_h, i \vec{p}_1)^2}{(i\vec{p}_1)^2}(-iE_1)+i \Lambda_{17(2)}  \frac{(-i \vec{p}_3, i \vec{p}_1)^2}{(i\vec{p}_1)^2}(-iE_1)\nonumber\\
    &+i \Lambda_{17(2)}  \frac{(-i \vec{p}_h, -i \vec{p}_3)^2}{(-i\vec{p}_3)^2}(iE_3)+i \Lambda_{17(2)}  \frac{(i \vec{p}_1, -i \vec{p}_3)^2}{(-i\vec{p}_3)^2}(iE_3).
\end{align*}

8) $\Lambda_{*(1)}\tilde{\mathcal{R}} \Big( \big(\partial^2 \tilde{\mathcal{R}} \big)^2-\big(\partial_i \partial_j \tilde{\mathcal{R}} \big)^2\Big)$ terms give: 
\begin{align*}
     &V_{*} = 2i\Lambda_{*(1)}(i\vec{p}_1)^2(-i\vec{p}_3)^2 +  2i\Lambda_{*(1)}(-i\vec{p}_3)^2(-i\vec{p}_h)^2+  2i\Lambda_{*(1)}(i\vec{p}_1)^2(-i\vec{p}_h)^2 \nonumber\\
     &- 2i\Lambda_{*(1)}(i \vec{p}_1, -i \vec{p}_3)^2 -2i\Lambda_{*(1)}(i \vec{p}_1, -i \vec{p}_h)^2  -2i\Lambda_{*(1)}(-i \vec{p}_3, -i \vec{p}_h)^2.
\end{align*}

Finally, the full expression for the first vertex with $\vec{p}_1$ and  $\vec{p}_3$  momenta from Fig.\ref{fig:channels} is given by
\begin{equation*}
    V_{p_1,p_3}= V_3 + V_4 +V_5 +V_6 + V_{13}+ V_{16}+ V_{17} + V_{*},
\end{equation*}
and all corresponding substitutions were made in \textit{Wolfram Mathematica}. The intermediate formula for $V_{p_1,p_3}$ is quite cumbersome, so we do not write the explicit expression here. 

In the same manner one can obtain the expressions for the second vertex $V_{p_2,p_4}$ (bottom vertex with $\vec{p}_2$ and $\vec{p}_4$ momenta) from the center diagram in Fig. \ref{fig:channels}. Actually, this  $V_{p_2,p_4}$ vertex can be easily obtained with the use of the following change of variables in final expression for $V_{p_1,p_3}$ together with the conservation laws \eqref{conservation}: $\vec{p}_1 \to \vec{p}_2 \to -\vec{p}_1$,  $\vec{p}_3 \to \vec{p}_4 \to - \vec{p}_3$, as well as $\vec{p}_h \to - \vec{p}_h$.

Both $V_{p_1,p_3}$ and $V_{p_2,p_4}$ contains $\Lambda_3-\Lambda_{6}, \Lambda_{*}, \Lambda_{13}, \Lambda_{16}, \Lambda_{17}$ coefficients, which are listed in Appendix \ref{app:lambdas} and different combinations of energies $E_1$ and $E_3$, momenta $\vec{p}_1$ and $\vec{p}_3$, and their scalar products together with $\vec{p}_h$. The needed combinations can be found with the use of conservation laws \eqref{conservation} and given by
\begin{equation*}
    (\vec{p}_1,\vec{p}_3) = \frac{E^2}{4c_S^2}\text{cos}\;\theta,
\end{equation*}
where $\theta$ is the angle between $\vec{p}_1$ and $\vec{p}_3$ vectors;
next
\begin{equation*}
    p_h^2 = p_1^2 + p_3^2 - 2p_1p_3\;\text{cos}\;\theta = \frac{E^2}{2c_S^2}(1-\text{cos}\;\theta),
\end{equation*}
and
\begin{equation*}
    (\vec{p}_1,\vec{p}_h) = p_1^2 - p_1p_3\;\text{cos}\;\theta = \frac{E^2}{4c_S^2}(1-\text{cos}\;\theta).
\end{equation*}
Substituting all these expressions together with \eqref{cs}, \eqref{Gs_cs_eps} into the formula for $t$-channel matrix element
\begin{align*}
    i M_t = \frac{i}{-p_h^2} V_{p_1,p_3} V_{p_2,p_4}
\end{align*}
one can arrive to \eqref{t-channel-acc}. 

Finally, as it was mentioned in the Section \ref{subsec:diagram}, one can obtain $u$-channel element \eqref{u-channel-acc} using the expression for $t$-channel \eqref{t-channel-acc} and performing the change $\text{cos}\;\theta \to - \text{cos}\;\theta$ there.

\section{Subtlety in the calculation of $s$-matrix element}
\label{app:s_ch}
\numberwithin{equation}{section}

\setcounter{equation}{0}

In this Appendix we discuss a subtlety, which arises in calculations for $s$-channel matrix element \eqref{s-channel-acc}. Formulas for $t$-  and $u$-channels can be obtained in a quite straightforward way and it was discussed in the previous Appendix. We remind, that we consider only such vertices in the matrix element which involve only $\Lambda_3-\Lambda_{6}, \Lambda_{*}, \Lambda_{13}, \Lambda_{16}, \Lambda_{17}$ couplings, since these terms provide the strongest naive constraints \eqref{str1} and \eqref{str2}, i.e. contributions from other terms are suppressed with $\epsilon$. 

The mentioned subtlety in $s$-channel is related to the terms with $\Lambda_{13}$, $\Lambda_{16}$, and $\Lambda_{17}$ couplings which involve $\psi = \partial^{-2}\dot{\mathcal{R}}$ in the cubic action \eqref{cubic}. Recalling that the momentum of propagator equals to zero for the $s$-channel (see conservation law for momenta \eqref{law_momenta}), we consider the following terms firstly
\begin{equation}
\label{app:l13}
    \Lambda_{13(1)} \partial^2 \tilde{\mathcal{R}}\partial_i \tilde{\mathcal{R}}\partial_i \partial^{-2} \tilde{\mathcal{R}} + \Lambda_{13(2)}\partial^2 \tilde{\mathcal{R}}\partial_i \tilde{\mathcal{R}}\partial_i \tilde{\psi}.
\end{equation}
One can easily see, that we get a $\frac{1}{0^2}$ factor as $\partial^{-2}$ acting on the propagator. To deal with such contributions, let us introduce a new parameter $\vec{\eta}\to 0$, which satisfies $\vec{p}_{1,2}\perp\vec{\eta}$, so we change the center-of-mass frame to a new frame with 
\begin{align*}	
\vec{p}_{1}\;'+\vec{p}_2\;' = \vec{\eta},
\end{align*}
where
\begin{subequations}
\label{app:eta_laws}
\begin{equation}
    \vec{p}\;'_{1,2} \to \vec{p}_{1,2} + \frac{\vec{\eta}}{2}.
\end{equation}
 Next, we find
\begin{align*}
    &(p_{1,2}')^2 = p_{1,2}^2 +\frac{(\vec{\eta}\;)^2}{4},\\ &(\vec{p}_{1}\;',\vec{p}_2\;') = (\vec{p}_{1},\vec{p}_2)+\frac{(\vec{\eta}\;)^2}{4},
\end{align*}
as well as
\begin{align}	E_1'=E_1\sqrt{1+\frac{c_S^2(\vec{\eta}\;)^2}{4E_1^2}}\approx E_1\Big(1+\frac{c_S^2(\vec{\eta}\;)^2}{8E_1^2}\Big).
\end{align}
\end{subequations} 
\begin{figure}[H]
    \centering
\includegraphics[scale=0.3]{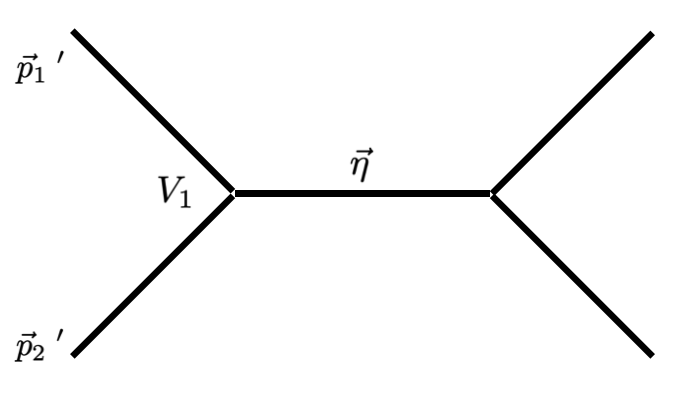}
    \caption{Tree level diagram for $2\to 2$ process: $s$-channel.}
    \label{fig:app_s}
\end{figure}
Thus, considering the left vertex from the diagram in Fig.\ref{fig:app_s} we write a related expression for the vertex connected with \eqref{app:l13} terms 
\begin{align*}
    &i\Lambda_{13(1)} (i\vec{p}\;'_1)^2 \frac{(i\vec{p}\;'_2,-i\vec{\eta}\;)}{(-i\vec{\eta}\;)^2} + i\Lambda_{13(1)} (i\vec{p}\;'_2)^2 \frac{(i\vec{p}\;'_1,-i\vec{\eta}\;)}{(-i\vec{\eta}\;)^2}\\
    &+i\Lambda_{13(2)} (i\vec{p}\;'_1)^2 \frac{(i\vec{p}\;'_2,-i\vec{\eta}\;)}{(-i\vec{\eta}\;)^2} (iE\;') + i\Lambda_{13(2)} (i\vec{p}\;'_2)^2 \frac{(i\vec{p}\;'_1,-i\vec{\eta}\;)}{(-i\vec{\eta}\;)^2} (iE').
\end{align*}
The same ``trick'' should be done for $\Lambda_{16}$ and $\Lambda_{17}$ terms. The final result with all contributions from $\Lambda_{13}$, $\Lambda_{16}$, and $\Lambda_{17}$ for the left vertex in Fig.~\ref{fig:app_s} reads
\begin{align}
\label{app:V1}
    &V_1 = \frac{1}{2} \Big(i \Lambda_{13(1)} \vec{p}_1\;^2 - \Lambda_{13(2)}  \vec{p}_1\;^2 E + i \Lambda_{13(1)} \vec{p}_2\;^2 - \Lambda_{13(2)} \vec{p}_2\;^2 E \Big)\nonumber\\
    &+\frac{|\vec{\eta}\;|^2}{2} \Big(-i\Lambda_{16 (1)} + \Lambda_{16 (2)}E - \frac{1}{2}\Lambda_{16 (3)}E_1 -\frac{1}{2}\Lambda_{16 (3)}E_2 -\frac{i}{2}\Lambda_{16 (4)}E_1 E - \frac{i}{2}\Lambda_{16 (4)}E_2 E \nonumber\\
    &-i \Lambda_{17 (1)} + \Lambda_{17 (2)} E\Big)\nonumber\\
    & = \frac{1}{2} \Big(i \Lambda_{13(1)} \vec{p}_1\;^2 - \Lambda_{13(2)}  \vec{p}_1\;^2 E + i \Lambda_{13(1)} \vec{p}_2\;^2 - \Lambda_{13(2)} \vec{p}_2\;^2 E \Big),
\end{align}
where in the last equality we take $\vec{\eta} = 0$ and also we substitute~\eqref{app:eta_laws}. Here we use $(i \vec{p}\;'_1, -i \vec{\eta}\;) = (\vec{p}_1 + \frac{\vec{\eta}}{2},  \vec{\eta}\;) = \frac{|\vec{\eta}\;|^2}{2}$ as well. Using the same logic one can consider the second right vertex with outcoming particles in Fig.~\ref{fig:app_s} and find out that similar contribution proportional to $\vec{\eta}$ vanishes in the same way as in \eqref{app:V1}. 

This concludes our discussion related to a subtlety coming from $\psi$ factor. We note once again, that $t$- and $u$-channels do not suffer from this problem, since the propagator's momentum is not zero in these cases.

\end{document}